# The two-dimensional hydrogen atom in
# The momentum representation

**M. Hage-Hassan**
Université Libanaise, Faculté des Sciences Section (1)
Hadath-Beyrouth

**Abstract**

The analytic expression of the momentum representation in terms of the associated Legendre function is determined by a direct integration of Fourier transform of the wave function of coordinates using the Levi-Civita transformation $R^2 \to R^2$ and the generating function method. A new generating function for the associated Legendre functions are obtained.

## 1. Introduction

The problem of the hydrogen atom in momentum space has been reformulated by Fock [1] which led to an integral form of the Schrödinger equation. This equation is solved by projecting the three-dimensional momentum space onto the surface of a four-dimensional sphere and the eigenfunctions are then expanded in terms of spherical harmonics. The applications of Fock's method to various problems of quantum mechanics are very extensive [2-4]. Recently, Fock's stereographic projection method had been applied to the theory of anisotropic excites and the two dimensional hydrogen atom. It had physical realization [5-7] and many papers treated this case [5-8].

In a previous paper [9] we presented a new and elementary method for the determination of the analytic expression for the wave function in momentum space in the third dimension space by a direct integration of the Fourier transform of the wave function of coordinates. We construct first, the generating function of the basis of coordinates assuming that the energy is a constant. Second, we perform the Fourier transform using Hurwitz transformations [10-11], or quadratic transformations to obtain the generating function for the momentum space. Finally, the wave function in the momentum space is derived from the development of the last function and the energy is replaced by its value.

The presented paper is part two of our previous work [9], we deal with the same method for the two dimensional hydrogen atom and we derive the wave function in the momentum representation. It is important to note that the calculation in the case of 2-dimensions is particular and differs from the treatment of 3-dimensions and from Fock's stereographic projection method.

This paper is organized as follows. In section two, we derive the wave function of the two dimensional hydrogen atom. In section three, we construct the generating function for the basis of the hydrogen atom and are also devoted to the presentation of the transformation $R^2 \to R^2$. In sections four and five, we derive the wave functions of the hydrogen atom in momentum space. In section six, we derive the last wave functions in terms of spherical harmonics. Finally in section seven, we acquire the new generating function for the associated Legendre functions.

## 2. The wave function of two dimensions hydrogen atom

The Schrödinger equation of the two dimensional hydrogen atom is:

$$H\Psi(\rho,\varphi) = -\frac{\hbar^2}{2\mu}\Delta - \frac{e^2}{\rho}\Psi(\rho,\varphi) \tag{2.1}$$

To solve the equation (2.1) we apply the method of separation of variables

$$\Psi(\rho,\varphi) = R(\rho)\Phi(\varphi) \tag{2.2}$$

$(\rho,\varphi)$ are the polar coordinates.

Introducing a separation constant, $m^2$, we can obtain the angular equation

$$\frac{d^2}{d\varphi^2}+m^2\varphi = 0 \tag{2.3}$$

The solution is:

$$\Phi(\varphi) = \frac{1}{\sqrt{2\pi}} e^{im\varphi} \tag{2.4}$$

The corresponding radial equation is:

$$-\frac{\hbar^2}{2\mu}[-\frac{1}{\rho}\frac{\partial}{\partial\rho}(\rho\frac{\partial}{\partial\rho})+\frac{1}{\rho^2}m^2]-\frac{e^2}{\rho}]R = ER \tag{2.5}$$

For simplicity, we choose the Gaussian units $\hbar = 2\mu = e^2/2$
and for negative energy we place $E = -q_0^2$,

We find the equation [12]
$$\frac{d^2R}{d\rho^2}+\frac{1}{\rho}\frac{dR}{d\rho}+(\frac{2}{\rho}-q_0^2-\frac{m^2}{\rho^2})R = 0 \tag{2.6}$$

We make the substitution $R(\rho) = (v)^{|m|}e^{-v/2}F(v)$ and $v = 2q_0\rho$
This leads us to the equation:

$$v\frac{d^2F}{dv^2}+(2|m|+1-v)\frac{dF}{dv}+(\frac{1}{q_0}-2|m|-\frac{1}{2})F = 0 \tag{2.7}$$

This is the confluent hypergeometric equation:
$$vy''+(2|m|+1-v)y'+ny = 0$$

This has two linearly independent solutions. If we choose the solution which is regular at the origin, then this becomes a polynomial of finite degree for $q_0 = (n+1/2)^{-1}$
and n = 0, 1, 2, . . .

The solutions of the equation (2.7) are the associated Laguerre polynomials $L_{n-|m|}^{2|m|}(v)$ [14] .We can now write the real-space wave function in the form:

$$\Psi_{nm}(\rho,\varphi) = N_{nm}(v)^{|m|}e^{-v/2}L_{n-|m|}^{2|m|}(v)e^{im\varphi} \qquad (2.8)$$

Using the orthogonal relations of Laguerre polynomials and the functional relation [14]

$$vL_n^\alpha(v) = (2n+\alpha+1)L_n^\alpha(v) - (n+1)L_{n+1}^\alpha(v) - (n+\alpha)L_{n-1}^\alpha(v)$$

We obtain the normalized wave functions:

$$\Psi_{nm}(\rho,\varphi) = N_{nm}(v)^{|m|}e^{-v/2}L_{n-|m|}^{2|m|}(v)e^{im\varphi} \qquad (2.9)$$

With
$$N_{nm} = \sqrt{\frac{q_0^3(n-|m|)!}{\pi(n+|m|)!}}$$

## 3. Generating function for the basis of two dimensions hydrogen atom and the quadratic transformation $R^2 \to R^2$

We construct the generating function for the basis of the hydrogen atom and we define the transformation $R^2 \to R^2$

### 3.1 The generating function of Laguerre polynomial $L_{n-|m|}^{2|m|}(v)$

The generating function of Laguerre polynomial is:

$$\sum_{n=0}^{\infty} z^n L_n^{(r)}(v) = \frac{1}{(1-z)^{r+1}} e^{-\frac{z}{1-z}v} \qquad (3.1)$$

From the property
$$\frac{d}{dv}L_n^{(\alpha)}(v) = -L_{n-1}^{(\alpha+1)}(v)$$

We deduce

$$\sum_{n=0}^{\infty} z^n L_{n-|m|}^{2|m|}(v) = \frac{(z)^{|m|}}{(1-z)^{2|m|+1}} \exp(-\frac{zv}{(1-z)}) \qquad (3.2)$$

### 3.2 The generating function
We assume that $q_0$ is a constant. We write

$$G(z,t,\vec{p}) = \sum_{nm} N_{nm}^{-1} z^n \frac{t^{|m|}}{|m|!}\Psi_{nm}(\rho,\varphi) =$$

$$e^{-q_0\rho}\sum_{nm} z^n L_{n-|m|}^{2|m|}(2q_0\rho)\frac{t^{|m|}}{|m|!}(2q_0\rho)^{|m|}e^{im\varphi} =$$

$$\frac{1}{(1-z)}e^{-q_0\rho}\exp(-\frac{z(2q_0\rho)}{(1-z)} + \frac{2tzq_0(x\pm iy)}{(1-z)^2} \qquad (3.3)$$

## 3.3 The quadratic transformation $R^2 \to R^2$

The well known Levi-Civita, $R^2 \to R^2$ transformation is
$$x = u_1^2 - u_2^2, \quad y = 2u_1 u_2 \tag{3.4}$$

## 3.4 The volume element

We consider the transformation $(u_1, u_2) \to (\rho, \varphi)$

With $\quad 0 \leq \varphi \leq 2\pi, \, 0 \leq \rho \leq \infty, \, -\infty \leq u_i \leq +\infty, i = 1,2.$

The calculation of the Jacobian gives $|J| = 4u^2$ and $d^2\vec{r} = r dr d\varphi$

Furthermore $\quad dxdy = \rho d\rho d\varphi = 4|J|d^2\vec{u}$

And $\quad 4u^2 d\vec{u} = d^2\vec{r}$

Therefore $\quad \int f(x,y) d^2\vec{r} = 4\int f(x(u), y(u)) u^2 d^2\vec{u} \tag{3.5}$

## 4. The generating function of the hydrogen atom in momentum space

We first write the Fourier transform in the representation (u) then we perform the integration thus, we obtain the generating function in the momentum representation.

### 4.1 The generating function in {u} representation

The wave function of the hydrogen atom in momentum space is:
$$\psi_{nm}(\vec{p}) = \frac{1}{2\pi} \int e^{-i\vec{p}\cdot\vec{r}} \psi_{nm}(\vec{r}) d^2\vec{r} \tag{4.1}$$

We denote the generating function by $G(z, t, \vec{p})$ in the representation $\{u\}$.

With $\quad G(z, t, \vec{p}) = \sum_{nm} N_{nm}^{-1} z^n \frac{t^{|m|}}{|m|!} \Psi_{nm}(\vec{p}) \tag{4.5}$

To calculate this expression (4.1) we must use the (u) representation and the formula (3.5):
$$\psi_{nm}(\vec{p}) = \frac{2}{\pi} \int e^{-i\vec{p}\cdot\vec{r}} \psi_{nm}(\vec{r}) u^2 d^2\vec{u} \tag{4.2}$$

in the above expression $\psi_{nlm}(\vec{p})$ contains the term $u^2$ for which we must consider a new generating function:
$$G(z, t, \vec{p}, \beta) = \frac{2}{\pi(1-z)} \int e^{-i\vec{p}\cdot\vec{r} - q_0\rho - \beta u^2} \exp\left(-\frac{z(2q_0\rho)}{(1-z)} + \frac{tzq_0(x \pm iy)}{(1-z)^2}\right) d^2\vec{u} \tag{4.3}$$

We assume that $\beta \geq 0$ as to eliminate the problem of convergence.

We then write $\quad [-\frac{\partial}{\partial \beta} G(z, t, \vec{p}, \beta)]\big|_{\beta=0} = G(z, t, \vec{p}) \tag{4.4}$

We chose m ≥0 for the following procedures because the calculations are not affected by this choice.

## 4.2 The generating function of momentum-space

We can do the integration of (4.3) by a direct calculation with the variables (u) using the integral $I = \int_E e^{-P} dx_1 dx_2 dx_3 ... dx_n$ with $P = \sum_{ij=1}^{n} a_{ij} x_i x_j$

The well known solution is $I = (\pi)^{\frac{n}{2}} / \sqrt{a}$, $a = \det(a_{ij})$

We have
$$\int e^{-i\vec{p}\cdot\vec{r}} \exp(-(q_0+\beta)\rho - \frac{z(2q_0\rho)}{(1-z)} + \frac{2tzq_0(x+iy)}{(1-z)^2}) d^2\vec{u}$$

$$= \int e^{-i\vec{p}\cdot\vec{r}} \exp(-[\frac{(1+z)(q_0)}{(1-z)} + \beta]\rho + \frac{2tzq_0(x+iy)}{(1-z)^2}) d^2\vec{u}$$

and
$$-i\vec{p}\cdot\vec{r} = -ip_x(u_1^2 - u_2^2) - ip_y(2u_1 u_2)$$

$$-[\frac{(1+z)(q_0)}{(1-z)} + \beta](u_1^2 + u_2^2) + \frac{2tzq_0((u_1^2 - u_2^2) + i(2u_1 u_2))}{(1-z)^2}$$

$$\rho = u_1^2 + u_2^2$$

We obtain then

$$X = \begin{pmatrix} [\frac{(1+z)(q_0)}{(1-z)} + \beta] - (\frac{2tzq_0}{(1-z)^2} - ip_x) & -i\frac{2tzq_0}{(1-z)^2} + ip_y \\ -i\frac{2tzq_0}{(1-z)^2} + ip_y & [\frac{(1+z)(q_0)}{(1-z)} + \beta] + (\frac{2tzq_0}{(1-z)^2} - ip_x) \end{pmatrix}$$

The $\det(X) = \frac{1}{(1-z)^2}\{[(1+z)q_0 + \beta(1-z)]^2 + \vec{p}^2(1-z)^2 + 4itzq_0(p_x + ip_y)\}$

We then find the generating function $G(z,t,\vec{p},\beta)$

$$G(z,t,\vec{p},\beta) = 2\frac{1}{\sqrt{\{[(1+z)q_0 + \beta(1-z)]^2 + \vec{p}^2(1-z)^2 + 4itzq_0(p_x + ip_y)\}}}$$

In applying the relation (4.4) we find the generating function $G(z,t,\vec{p})$

$$G(z,t,\vec{p}) = 2\frac{(1-z^2)q_0}{\{[(1+z)q_0]^2 + \vec{p}^2(1-z)^2 + 4itzq_0(p_x + ip_y)\}^{3/2}} \qquad (4.6)$$

## 5. The wave functions in momentum-space

The development of the function (4.6) gives us the wave functions of the hydrogen atom in momentum space. We derive the basis of momentum-space using the formula

$$[\frac{1}{n!}\frac{\partial^n}{\partial z^n}\frac{\partial^{|m|}}{\partial t^{|m|}} G(z,t,\vec{p})]\Big|_0 = N_{nm}^{-1} \Psi_{nm}(\vec{p}) \qquad (5.1)$$

In this level we must take $q_0 = 1/(n+1/2)$ as to execute the calculations proceed by the following steps:

1 - <u>Derivation with respect to t</u>

$$[\frac{\partial^{|m|}}{\partial t^{|m|}}G(z,t,\vec{p})]\Big|_0 = 2\frac{(-1)^{|m|}(3/2)_{|m|}(1-z^2)z^{|m|}[4iq_0(p_x+ip_y)]^m}{\{[(1+z)q_0]^2+\vec{p}^2(1-z)^2\}^{|m|+3/2}}$$

We obtain $(q_0^2(1+z))^2+(1-z)^2\vec{p}^2 = ((\vec{p}^2+q_0^2)-2z(\vec{p}^2-q_0^2)+z^2(\vec{p}^2+q_0^2))$

$$= (\vec{p}^2+q_0^2)[1-2zq+z^2], \text{ with } q = \left(\frac{\vec{p}^2-q_0^2}{\vec{p}^2+q_0^2}\right)$$

We deduce that

$$[\frac{\partial^m}{\partial t^m}G(z,t,\vec{p})]\Big|_0 = 2\frac{(-4i)^m(q_0)^{m+1}(3/2)_m}{(\vec{p}^2+q_0^2)^{|m|+3/2}}\frac{(1-z^2)z^{|m|}}{\{[1-2zq+z^2\}^{|m|+3/2}}[(p_x+ip_y)]^m \quad (5.2)$$

2- <u>Derivation with respect to z</u>

Using the familiar formula for the generating function of Gegenbauer polynomials

$$(1-2qz+z^2)^{-\alpha} = \sum_{m=0}^{\infty} z^m C_m^\alpha(q)$$

We write

$$\frac{(1-z^2)z^{|m|}}{[1-2zq+z^2]^{|m|+3/2}} = \sum_{k=0}^{\infty}(1-z^2)z^{|m|}z^k C_k^{|m|+3/2}(q) \quad (5.3)$$

$$= \sum_n z^n \left[C_{n-|m|}^{|m|+3/2}(q) - C_{n-|m|-2}^{|m|+3/2}(q)\right] \quad (5.4)$$

With $k+|m|=n$, $k+|m|+2=n$ and $q_0=1/(n+1/2)$ therefore,

$$[\frac{1}{n!}\frac{\partial^n}{\partial z^n}\frac{\partial^{|m|}}{\partial t^{|m|}}G(z,t,\vec{p})]\Big|_0 = 2\frac{(-4i)^{|m|}(q_0)^{|m|+1}(3/2)_{|m|}}{(\vec{p}^2+q_0)^{|m|+3/2}} \times$$

$$[C_{n-|m|}^{|m|+3/2}(q) - C_{n-|m|-2}^{|m|+3/2}(q)][(p_x+ip_y)]^m \quad (5.5)$$

3- <u>The wave function in momentum space</u>

With the help of the recurrences formula [13]

$$(n+1/2)C_{n-|m|}^{|m|+1/2}(q) = (m+1/2)[C_{n-|m|}^{|m|+3/2}(q) - C_{n-|m|-2}^{|m|+3/2}(q)] \quad (5.6)$$

We find: $[\frac{1}{n!}\frac{\partial^n}{\partial z^n}\frac{\partial^{|m|}}{\partial t^{|m|}}G(z,t,\vec{p})]\Big|_0 = 2\frac{(2n+1)}{(2|m|+1)}\frac{(-4i)^{|m|}(q_0)^{|m|+1}(3/2)_{|m|}}{(\vec{p}^2+q_0^2)^{|m|+3/2}} \times$

$$C_{n-|m|}^{|m|+1/2}(q)[(p_x+ip_y)]^m \quad (5.6)$$

Comparing (5.4) and (5.1) gives us the wave function in momentum space:

$$\Psi_{nm}(\vec{p}) = N_{nm}\frac{(2n+1)}{(2|m|+1)}\frac{2(-4i)^{|m|}(q_0)^{|m|+1}(3/2)_m}{(\vec{p}^2+q_0)^{|m|+3/2}}C_{n-|m|}^{|m|+1/2}(q)[(p_x+ip_y)]^m \quad (5.7)$$

It is important to note that $C_{n-|m|}^{|m|+1/2}(q)$ is a particular function of Gegenbauer polynomials. This function has a connexion with the associated Legendre polynomials.

## 6. The wave functions in momentum space in terms of spherical harmonics

The connection of Gegenbauer polynomials with the associated Legendre polynomials is given by the formula [14]

$$C_{n-|m|}^{|m|+1/2}(t) = \frac{1}{(2|m|-1)!!} \frac{d^{|m|} P_n(t)}{dt^{|m|}} = (-1)^{|m|} \frac{(1-t^2)^{-|m|/2} |m|! 2^{|m|}}{(2|m|)!} P_n^{|m|}(t) \qquad (6.1)$$

We derive finally the wave functions in momentum space:

$$\Psi_{nm}(\vec{p}) = N_{nm} \frac{(n+1/2)}{(|m|+1/2)} \frac{2(-4i)^{|m|}(q_0)^{|m|+1}(3/2)_{|m|}}{(\vec{p}^2 + q_0^2)^{|m|+3/2}}$$

$$(-1)^m \frac{(1-q^2)^{-|m|/2} |m|! 2^{|m|}}{(2|m|)!} P_n^{|m|}(q)[(p_x + ip_y)]^m \qquad (6.2)$$

But $(1-q^2)^{|m|/2} = \left(1 - \left(\frac{(\vec{p}^2 - q_0^2)}{(\vec{p}^2 + q_0^2)}\right)^2\right)^{|m|/2} = 2^{|m|} [\vec{p}^2]^{|m|/2} [q_0^2]^{|m|/2} / (\vec{p}^2 + q_0^2)^{|m|}$ (6.3)

And $$(3/2)_{|m|} = \frac{(2|m|+1)!}{4^{|m|} |m|!} \qquad (6.4)$$

Replacing (6.3) and (6.4) in (6.2) we find the wave functions in momentum space in terms of spherical harmonics:

$$\Psi_{nm}(\vec{p}) = (i)^{|m|} \sqrt{\frac{2(n-|m|)!}{\pi(n+|m|)!}} \left(\frac{2q_0}{\vec{p}^2 + q_0^2}\right)^{3/2} P_n^{|m|}(q) e^{im\varphi_p} \qquad (6.5)$$

It is clear that we obtain by an elementary method not only the wave function in momentum representation [5] but also the phase factor and the calculation in the case of 2-dimensions. It is particular and differs from the treatment of 3-dimensions.

## 7. The new generating function for the associated Legendre functions

We start from the expressions (5.3) and (5.4). We write that:

$$\frac{(1-z^2)z^{|m|}}{[1-2zt+z^2]^{|m|+3/2}} = \sum_{n=1}^{\infty} z^n \left[C_{n-|m|}^{|m|+3/2}(t) - C_{n-|m|-2}^{|m|+3/2}(t)\right]$$

By using expression (5.6) we find

$$\frac{(1-z^2)z^{|m|}}{[1-2zt+z^2]^{|m|+3/2}} = \sum_{n=1}^{\infty} z^n \frac{(2n+1)}{(2m+1)} C_{n-|m|}^{|m+1/2|}(t)$$

By connecting Gegenbauer polynomials with the associated Legendre polynomials we find a new generating function of these last polynomials

$$\frac{(1-t^2)^{|m|/2}(1-z^2)z^{|m|}}{[1-2zt+z^2]^{|m|+3/2}} = \sum_{n=1}^{\infty} (-1)^{|m|} z^n \frac{(2n+1)}{(2m+1)!!} P_n^{|m|}(t)$$

Finally, we note that the Fourier transformation for the hydrogen atom transforms the radial part into angular which explains the Fock transformation.
We also observe that our method for the 2-dimensional case is important not only for the physical applications [5-7] but also for educational applications because our result can also be an introduction to Octonion algebra and of course to Fock's transformations.